\documentstyle[buckow]{article}
\let\du=\du                     

\def\a{\alpha}
\def\b{\beta}
\def\c{\chi}

\def\j{\psi}

\def\l{\lambda}
\def\m{\mu}

\def\o{\omega}
\def\p{\pi}

\def\t{\tau}

\def\x{\xi}
\def\z{\zeta}

\def\G{\Gamma}

\def\L{\Lambda}

\def\S{\Sigma}

\def\ve{\varepsilon}
\def\vf{\varphi}

\def\cf{{\cal F}}

\def\ch{{\cal H}}

\def\ck{{\cal K}}

\def\bo{{\raise-.5ex\hbox{\large$\Box$}}}               
\def\pa{\partial}                                       
\def\TH{{\raise.2ex\hbox{$\displaystyle \bigodot$}\mskip-4.7mu \llap H \;}}
\def\face{{\raise.2ex\hbox{$\displaystyle \bigodot$}\mskip-2.2mu \llap {$\ddot
        \smile$}}}                                      

\def\Bar#1{\overline{#1}}                       
\def\sbar#1{\stackrel{*}{\Bar{#1}}}             
\def\VEV#1{\left\langle #1\right\rangle}        
\def\abs#1{\left| #1\right|}                    
\def\leftrightarrowfill{$\mathsurround=0pt \mathord\leftarrow \mkern-6mu
        \cleaders\hbox{$\mkern-2mu \mathord- \mkern-2mu$}\hfill
        \mkern-6mu \mathord\rightarrow$}
\def\dvec#1{\vbox{\ialign{##\crcr
        \leftrightarrowfill\crcr\noalign{\kern-1pt\nointerlineskip}
        $\hfil\displaystyle{#1}\hfil$\crcr}}}           
\def\dt#1{{\buildrel {\hbox{\LARGE .}} \over {#1}}}     

\def\frac#1#2{{\textstyle{#1\over\vphantom2\smash{\raise.20ex
        \hbox{$\scriptstyle{#2}$}}}}}                   
\def\sfrac#1#2{{\vphantom1\smash{\lower.5ex\hbox{\small$#1$}}\over
        \vphantom1\smash{\raise.4ex\hbox{\small$#2$}}}} 
\def\bfrac#1#2{{\vphantom1\smash{\lower.5ex\hbox{$#1$}}\over
        \vphantom1\smash{\raise.3ex\hbox{$#2$}}}}       
\def\afrac#1#2{{\vphantom1\smash{\lower.5ex\hbox{$#1$}}\over#2}}    

\def\[{\lfloor{\hskip 0.35pt}\!\!\!\lceil}
\def\]{\rfloor{\hskip 0.35pt}\!\!\!\rceil}

\def\du#1#2{_{#1}{}^{#2}}
\def\ud#1#2{^{#1}{}_{#2}}

\def\ha{{\fracmm12}}

\def\fracmm#1#2{{{#1}\over{#2}}}

\def\low#1{{\raise -3pt\hbox{${\hskip 0.75pt}\!_{#1}$}}}

\def\Dot#1{\buildrel{_{_{\hskip 0.01in}\bullet}}\over{#1}}
\def\dt#1{\Dot{#1}}

\newskip\humongous \humongous=0pt plus 1000pt minus 1000pt

\newif\ifdtup

\begin {document}
\def\titleline{
The hypermultiplet low-energy effective action,
\newtitleline
N=2 supersymmetry breaking and confinement
}
\def\authors{
Sergei V. Ketov  
}
\def\addresses{
Institut f\"ur Theoretische Physik, Universit\"at Hannover, \\
Appelstrasse 2, D--30167 Hannover\\
}
\def\abstracttext{
Some exact solutions to the hypermultiplet low-energy effective action in 
$N=2$ supersymmetric four-dimensional gauge field theories with massive 
`quark' hypermultiplets are discussed. The need for a spontaneous $N=2$
supersymmetry breaking is emphasized, because of its possible relevance in 
the search for an ultimate theoretical solution to the confinement problem.
}

\noindent ITP--UH--34/97 \hfill hep-th/9712152 

\makefront

\vglue.3in

\begin{center}
\noindent
{\it Talk given at the 31st International Ahrenshoop Symposium on the
Theory of Elementary Particles, \\
Buckow, Brandenburg, Germany, 2--6 September 1997 \\
 and \\
the String Workshop in Wittenberg, Germany, 14--15 November 1997}
\end{center}

\newpage

\section{Introduction}

Despite of remarkable recent advances in string duality and brane technology, 
our knowledge about the non-perturbative string theory (= M-theory) is still
very much dependent upon our understanding of non-perturbative quantum field
theory like QCD. At high energies the QCD is well described by a perturbation
theory because of its asymptotic freedom, in a good agreement with well-known 
experimental data about deep inelastic scattering and jet production. However,
at low energies, the QCD vacuum is essentially non-perturbative, so that some
of the most obvious experimental facts about strong interactions, e.g., the 
confinement of quarks inside hadrons, are still waiting for an ultimate 
theoretical solution. On the theoretical side, the quantum generating 
functional (or the effective action) of a non-abelian gauge field theory 
should be defined in practical terms, which would allow one to get a 
non-perturbative solution to the theory. Unfortunately, the corresponding
path integral is usually defined in many ways beyond the perturbation  
theory (e.g., lattice field theory, instantons, duality), which makes getting
an exact solution to be extremely difficult, if ever possible.  

Therefore, it seems to be quite natural to take advantage of the existence of
{\it exact} solutions to the low-energy effective action in certain $N=2$
supersymmetric gauge field theories since the remarkable discovery of Seiberg
and Witten \cite{sw}, and apply them to the old problem of color confinement 
in QCD. In fact, it was one of the main motivations in the original work 
\cite{sw}.

The most attractive mechanism for color confinement is known to be the dual 
Meissner effect or the dual (Type II) superconductivity \cite{tman}. It
takes just three steps to connect an ordinary BCS superconductor to the 
simplest Seiberg-Witten model in quantum field theory: first, define a 
relativistic version of the superconductor, known as the (abelian) Higgs model
in field theory, second, introduce a non-abelian version of the Higgs model,
known as the Georgi-Glashow model, and, third, $N=2$ supersymmetrize the 
Georgi-Glashow model in order to get the Seiberg-Witten model \cite{sw}. Since
the t'Hooft-Polyakov monopole of the Georgi-Glashow model belongs to a (HP)
hypermultiplet in its $N=2$ supersymmetric (Seiberg-Witten) generalisation,
it is quite natural to explain confinement as the result of a monopole 
condensation (dual Higgs effect), i.e. a non-vanishing vacuum expectation 
value for the magnetically charged (dual Higgs) scalars belonging to the HP 
hypermultiplet. 

In fact, the exact solutions to the low-energy effective action in quantum
gauge field theories are only available in $N=2$ supersymmetry, and neither 
in $N=1$ supersymmetry nor in the bosonic QCD. Hence, on the one side, it is 
the $N=2$ supersymmetry that crucially simplifies an evaluation of the 
low-energy effective action. However, on the other side, it is the same $N=2$ 
supersymmetry that is obviously incompatible with phenomenology e.g., because 
of equal masses of bosons and fermions inside $N=2$ supermultiplets (it also
applies to any $N\geq 1$ supersymmetry), and the non-chiral nature of $N=2$
supersymmetry (e.g. `quarks' then appear in real representations of the gauge 
group). Therefore, if we believe in the $N=2$ supersymmetry, we should find a 
way of judicious $N=2$ supersymmetry breaking. 

The $N=2$ supersymmetry can be broken either softly or spontaneously, if one
wants to preserve the benefits of its presence (e.g. for the full control
over the low-energy effective action) at high energies. As regards the 
{\it gauge} low-energy effective action, the information about it in the 
Seiberg-Witten approach is encoded in terms of holomorphic functions defined 
over the quantum moduli space whose modular group is identified with the 
duality group, while the functions themselves can be calculated exactly. In 
the $N=2$ supersymmetric QCD, one has to add `quark' hypermultiplets, which 
have some bare (BPS) masses, flavour and color, i.e. belong to the fundamental
representation of the gauge group. In the full theory, one expects an
appearance of additional (e.g. magnetically charged) degrees of freedom, to be
described by some effective action via strong-weak coupling duality and 
depending upon the (Coulomb, Higgs or confinement) branch under consideration.
The full low-energy effective action in the $N=2$ super-QCD is given by a sum 
of the gauge and the hypermuliplet parts. 

We would like to find a vacuum solution to the full $N=2$ supersymmetric 
low-energy effective action, which would break supersymmetry due to the 
non-vanishing vacuum expectation value of a magnetically charged (Higgs) 
scalar, similarly to that in ref.~\cite{sw}. The same dual Higgs mechanism 
may also be responsible for the chiral symmetry breaking and the appearance of
the pion effective Lagrangian if the dual Higgs field has flavor charges also
~\cite{sw}. In fact, Seiberg and Witten used a mass term for the $N=1$ chiral
multiplet, which is a part of the $N=2$ vector multiplet, in order to 
{\it softly} break $N=2$ supersymmetry to $N=1$ supersymmetry. As a result, 
they found a non-trivial vacuum solution with a monopole condensation and, 
hence, a confinement. The weak point of their approach is an {\it ad hoc}
assumption about the existence of the mass gap, i.e. the mass term itself. It
would be nice to derive the mass gap from the fundamental theory instead of
postulating it. The $N=2$ supersymmetry may be useful here since it severely
constrains all possible ways of its soft (or spontaneous) breaking.

The {\it soft} breaking of $N=2$ supersymmetry is a very practical approach to
analyse the consequences of the Seiberg-Witten exact solution towards its 
possible phenomenological applications like a derivation of the pion 
lagrangian or the confinement problem in QCD. The general analysis of all 
possible soft $N=2$ supersymmetry breaking patterns in the $N=2$ 
supersymmetric QCD was recently given by Alvarez-Gaum\'e, Mari\~no and Zamora 
in ref.~\cite{amz}. Though being quite pragmatic, the soft susy breaking has, 
however, a limited predictive power and too many free parameters. Hence, it 
makes sense to search for the patterns of {\it spontaneous} $N=2$
supersymmetry breaking. In practice, this means finding a non-supersymmetric 
vacuum solution for the $N=2$ supersymmetric scalar potential at the level of
the low-energy effective action in $N=2$ gauge theories. Since the $N=2$
supersymmetry remains unbroken for any exact Seiberg-Witten solution in the
gauge sector, we should consider the induced (i.e. quantum generated) scalar
potentials in the hypermultiplet sector of an $N=2$ gauge theory. Moreover,
once we accepted $N=2$ supersymmetry in field theory, we can also take into 
account those brane configurations of the underlying M-theory that are relevant
for the four-dimensional $N=2$ supersymmetric effective physics in the limit 
$M_{\rm Planck}\to\infty$. The related brane-technology \cite{wi} can provide 
us with some additional insights into the non-perturbative field theory, as 
well as supply us with its geometrical interpretation. Since the relevant 
M-theory brane configurations with eight supercharges arise as the solitonic 
solutions to the effective equations of motion in the M-theory, their `soft' 
deformation, which breaks some more of the supersymmetries but still remains 
to be a solution to the M-theory effective equations of motion, should be 
interpreted as a spontaneous supersymmetry breaking 
(see an example in sect.~3).

In sect.~2 we analyse the general problem of constructing the low-energy
hypermultiplet effective action in $N=2$ rigid (global) supersymmetry, by 
using the $N=2$ harmonic superspace. In sect.~3 we give two simple (toy) 
examples of the non-trivial induced scalar potentials for a single matter 
hypermultiplet.  
  
\section{The hypermultiplet low-energy effective action}

There are only two basic $N=2$ supermultiplets (modulo classical duality
transformations) in the rigid $N=2$ supersymmetry (with the $SU(2)_A$ internal
symmetry): an $N=2$ vector multiplet and a hypermultiplet. The $N=2$ vector 
multiplet components (in a WZ-like gauge) are $(A,\l^i_{\a},V_{\m},D^{(ij)})$,
where $A$ is a complex Higgs scalar, $\l^i$ is a chiral spinor (`gaugino') 
$SU(2)_A$ doublet, $V_{\m}$ is a real gauge vector field, and $D^{(ij)}$ is an
auxiliary  $SU(2)_A$ scalar triplet $(i,j=1,2)$. Similarly, the on-shell 
physical 
components of the Fayet-Sohnius (FS)-type hypermultiplet are $(q^i,\j_{\a},
\bar{\j}_{\dt{\a}})$, where $q^i$ is a complex scalar $SU(2)_A$ doublet, and 
$\j$ is a Dirac spinor. There exists another (dual) Howe-Stelle-Townsend 
(HST)-type hypermultiplet, whose on-shell physical components are 
$(\o,\o^{(ij)},\c^i_{\a})$, where $\o$ is a real scalar, $\o^{(ij)}$ is a 
scalar $SU(2)_A$ triplet, and $\c^i$ is a chiral spinor (`quark') $SU(2)_A$ 
doublet.

The universal (i.e. most general and off-shell) and manifestly $N=2$ 
supersymmetric formulation of all $N=2$ supersymmetric four-dimensional
field theories is only possible in the $N=2$ harmonic superspace 
(HSS)~\cite{gikos} (see e.g, ref.~\cite{my} for a recent introduction). The
$N=2$ HSS coordinates include extra bosonic variables (called harmonics 
$u^{\pm}_i$), which parametrize the sphere $S^2\sim SU(2)/U(1)$, in addition
to the standard $N=2$ superspace coordinates. The harmonics play the role of
twistors or spectral parameters known in the theory of integrable systems. In
particular, an off-shell FS hypermultiplet in HSS is described by an analytic
superfield $q^+$ of the $U(1)$-charge $(+1)$, whereas the HST hypermultiplet
in HSS is described by a real analytic superfield $\o$ of vanishing $U(1)$ 
charge. An $N=2$ vector gauge multiplet is similarly described by an analytic 
HSS superfield $V^{++}$ of the $U(1)$-charge $(+2)$, which is introduced as a 
connection to the basic HSS harmonic derivative $D^{++}$ present in the 
kinetic terms of the hypermultiplet actions (see below). 

The power of $N=2$ superspace is clearly seen in the most general form of 
the $N=2$ gauge low-energy effective action in the Coulomb branch,
$$ \G_V[W,\bar{W}]=\int_{\rm chiral} \cf(W) +{\rm h.c.} + \int_{\rm full}
\ch(W,\bar{W}) + \ldots~,\eqno(1)$$
where the abelian field strength $W(V)$, which is a harmonic-independent, $N=2$
 chiral and gauge-invariant superfield, has been introduced. The leading term 
in eq.~(1) is given by the chiral $N=2$ superspace integral over a 
{\it holomorphic} function $\cf$ of $W$, with the latter being valued in the 
Cartan subalgebra of the gauge group. The Seiberg-Witten approach provides a 
solution to the holomorphic function $\cf$ in terms of the auxiliary Riemann 
surface $\S_{SW}$. It appears to be a solution to the particular 
Riemann-Hilbert problem of fixing a holomorphic multi-valued function $\cf$ by
its given monodromy and singularities. The number (and nature) of the 
singularities is the physical input: they are identified with the appearance 
of massless non-perturbative BPS-like physical states (dyons) like the 
t'Hooft-Polyakov magnetic monopole. The monodromies are supplied by 
perturbative renormalization-group $\b$-functions and S-duality. The 
next-to-leading-order term in eq.~(1) is given by the full $N=2$ superspace 
integral over a real function $\ch$ of $W$ and $\bar{W}$. Some partial results
about this function are known~\cite{next}. The dots in eq.~(1) stand
for higher-order terms containing the derivatives of $W$ and $\bar{W}$.

The most general form of the leading term in the hypermultiplet low-energy
effective action can be written down in the $N=2$ HSS as follows:
$$ \G_H[q^+,\sbar{q}{}^+;\o]=\int_{\rm analytic} \ck^{(+4)}(q^+,\sbar{q}{}^+;
\o;u^{\pm}_i)+\ldots~,\eqno(2)$$
where $\ck^{(+4)}$ is a function of the FS analytic superfield $q^+$, 
its conjugate $\sbar{q}{}^+$, the HST analytic superfield $\o$ and the 
harmonics $u^{\pm}_i$, with the overall $U(1)$-charge $(+4)$. The action (2) 
is supposed to be added to the kinetic hypermultiplet action whose analytic 
Lagrangian is quadratic in $q^+$ or $\o$, and of $U(1)$-charge $(+4)$. 
A free FS hypermultiplet action is given by
$$ S[q]=-\int d\z^{(-4)}du\,\sbar{q}{}^+D^{++}q^+~,\eqno(3)$$
whereas its minimal coupling to an $N=2$ gauge superfield reads
$$ S[q,V]= -\int d\z^{(-4)}du \,\sbar{q}{}^+(D^{++}+iV^{++})q^+~.
\eqno(4)$$
Similarly, a free action of the HST hypermultiplet is given by
$$S[\o]=-\frac{1}{2}\int d\z^{(-4)}du \,(D^{++}\o)^2~,\eqno(5)$$
and it is on-shell equivalent to the standard $N=2$ tensor (or linear)
multiplet action in the ordinary $N=2$ superspace \cite{my}.

The function $\ck$ is called the {\it hyper-K\"ahler potential}. In components,
it automatically leads to the $N=2$ supersymmetric non-linear sigma-model for
the scalars with a {\it hyper-K\"ahler} metric, just because of the $N=2$ 
supersymmetry by construction (see the examples in sect.~3). When being 
expanded in components, the first term in eq.~(1) also leads to the certain 
K\"ahler non-linear sigma-model in the Higgs sector $(A,\bar{A})$. The 
corresponding K\"ahler potential $K_{\cf}(A,\bar{A})$ is dictated by the 
holomorphic function $\cf$ as $K_{\cf}={\rm Im}[\bar{A}\cf'(A)]$, so that the 
function $\cf$ plays the role of a potential for this {\it special} K\"ahler 
(but not hyper-K\"ahler) geometry $K_{\cf}(A,\bar{A})$. As regards the 
hypermultiplet non-linear sigma-model of eqs.~(2)--(5), a relation between the
hyper-K\"ahler potential $\ck$ and the corresponding K\"ahler potential 
$K_{\ck}$ is much more involved. It is easy to see that the hyper-K\"ahler 
condition on a K\"ahler potential amounts to a non-linear (Monge-Ampere) 
partial differential equation. It is remarkable that the HSS approach allows 
one to get a formal 'solution' to any hyper-K\"ahler geometry in terms of an 
analytic scalar potential $\ck$. However, the real problem is now translated 
into finding the relation between $\ck$ and the corresponding K\"ahler 
potential (or metric) in components, whose determination amounts to solving 
infinitely many linear differential equations altogether, just in order to 
eliminate an infinite number of HSS auxiliary fields (sect.~3). 

The gauge-invariant functions $\cf(W)$ and $\ch(W,\bar{W})$ receive both 
perturbative and non-perturbative contributions,
$$ \cf= \cf_{\rm per.} + \cf_{\rm inst.}~,\qquad \ch=\ch_{\rm per.} +\ch_{\rm
non-per.}~,\eqno(6)$$
while the non-perturbative corrections to the holomorphic function $\cf$ are
entirely due to instantons. This is an important difference from the (bosonic)
non-perturbative QCD whose low-energy effective action is dominated by 
instanton-antiinstanton contributions.

It is remarkable that the perturbative contributions to the leading and 
subleading terms in the $N=2$ gauge effective action (1) come from the one 
loop only. As regards the leading holomorphic contribution, $N=2$ 
supersymmetry puts the trace of the energy-momentum tensor $T\du{\m}{\m}$ and 
the axial or chiral anomaly $\pa_{\m} j^{\m}_R$ of the abelian $R$-symmetry 
into one $N=2$ supermultiplet. The $T\du{\m}{\m}$ is essentially determined by
the perturbative renormalization group $\b$-function, $T\du{\m}{\m}\sim
\b(g)FF$, whereas the one-loop contribution to the chiral anomaly,
$\pa\cdot j_R\sim C_{\rm 1-loop}F{}^*F$, is known to saturate the exact
solution to the Wess-Zumino consistency condition for the same anomaly.
Hence, $\b_{\rm per.}(g)=\b_{\rm 1-loop}(g)$ by $N=2$ supersymmetry also.
Since the $\b_{\rm per.}(g)$ is effectively determined by the second
derivative of $\cf_{\rm per.}$, one concludes that 
$\cf_{\rm per.}=\cf_{\rm 1-loop}$ too. This simple component argument can be 
extended to a proof when using the $N=2$ HSS approach \cite{bko}. It than 
becomes clear that the non-vanishing central charges of the $N=2$ 
supersymmetry algebra are of crucial importance for the non-vanishing 
holomorphic contribution to the gauge effective action (1). 

Similarly, the BPS mass of a hypermultiplet can only come from the central 
charges since, otherwise, the number of the massive hypermultiplet components 
has to be increased. The most natural way to introduce central charges 
$(Z,\bar{Z})$ is to identify them with spontaneously broken $U(1)$ generators 
of dimensional reduction from six dimensions via the Scherk-Schwarz mechanism
\cite{ikz}. It naturally leads to the additional `connection' term in the 
four-dimensional harmonic derivative as
$$ {\cal D}^{++}=D^{++}+v^{++}~,\quad {\rm where}\quad
v^{++}=i(\theta^+\theta^+)\bar{Z}+i(\bar{\theta}^+\bar{\theta}^+)Z~.
\eqno(7)$$
Therefore, the $N=2$ central charges can be equally treated as a non-trivial 
$N=2$ gauge background, with the covariantly constant $N=2$ chiral superfield 
strength $\VEV{W}=Z$.

\section{Examples}

We are still far from presenting a convincing pattern of spontaneous $N=2$ 
supersymmetry breaking via the hypermultiplet low-energy effective action. 
Nevertheless, the examples that we already have, give some reasons for 
optimism. Our point here is quite simple: given non-trivial kinetic terms in 
the hypermultiplet low-energy effective action to be represented by the 
non-linear sigma-model, in a presence of non-vanishing central charges it 
leads to a non-trivial hypermultiplet scalar potential whose form is entirely 
determined by the hyper-K\"ahler metric of the kinetic terms and $N=2$ 
supersymmetry.

The first example of this interesting connection was given in ref.~\cite{ikz}.
Consider a single charged FS hypermultiplet $q^+$ in the Coulomb branch of the
$N=2$ gauge theory. As was shown in ref.~\cite{ikz}, it has a unique 
non-trivial self-interaction whose form in the $N=2$ HSS reads 
$$ LEEA[q^+]_{\rm Taub-NUT} = \int_{\rm analytic} \left[ \sbar{q}{}^+D^{++}q^+
+ \fracmm{\l}{2}(q^+)^2(\sbar{q}{}^+)^2\right]~,\eqno(8)$$
where the induced coupling constant $\l$ is given by
$$ \l=\fracmm{g^4}{\p^2}\left[ \fracmm{1}{m^2}\ln\left( 1+\fracmm{m^2}{\L^2}
\right) -\fracmm{1}{\L^2+m^2}\right]~,\eqno(9)$$
in terms of the gauge coupling constant $g$, the hypermultiplet BPS mass
$m^2=\abs{Z}^2$, and the IR-cutoff $\L$. When using the parametrization
$$ \left. q^+\right|_{\theta=0}=f^i(x)u^+_i\exp\left[ \l f^{(j}(x)\bar{f}^{k)}
(x)u^+_j u^-_k\right]~,\eqno(10)$$
the bosonic terms take the form of the one non-linear sigma-model,
$$ LEEA_{\rm bosonic}[f]=
\int d^4x\,\left\{ g_{ij}(f)\pa_mf^i\pa^mf^j +\bar{g}^{ij}
(f)\pa_m\bar{f}_i\pa^m\bar{f}_j +h\ud{i}{j}(f)\pa_mf^j\pa^m\bar{f}_i -V(f)
\right\}~,\eqno(11)$$
whose metric turns out to be that of Taub-NUT or a KK-monopole (modulo field 
redefinitions), whereas the induced scalar potential is~\cite{ikz}
$$V(f) =\abs{Z}^2\fracmm{f\bar{f}}{1+\l f\bar{f}}~~~.\eqno(12)$$

A non-trivial hypermultiplet self-interaction for a single neutral HST-type
$\o$-hypermultiplet can be non-perturbatively generated in the presence of 
non-vanishing constant $N=2$ Fayet-Iliopoulos (FI) term $\VEV{D^{(ij)}}\equiv
\x^{(ij)}=\ha(\vec{\t}\cdot\vec{\x})^{ij}$, where $\vec{\t}$ are Pauli 
matrices. The FI-term has a nice geometrical interpretation in the underlying 
ten-dimensional type-IIA superstring brane picture made out of two
solitonic 5-branes located at particular values of $\vec{w}=(x^7,x^8,x^9)$ and
some Dirichlet 4- and 6-branes, all having the four-dimensional spacetime
$(x^0,x^1,x^2,x^3)$ as the common macroscopic world-volume~\cite{wi}. The
values of $\vec{\x}$ can then be identified with the {\it angles\/} at which
the two 5-branes intersect, $\vec{\x}=\vec{w}_1-\vec{w}_2$, in the type-IIA
picture~\cite{my}. The three hidden dimensions $(\vec{w})$ are identified
by the requirements that they do not include the two hidden dimensions
$(x^4,x^5)$ already used to generate central charges in the effective
four-dimensional field theory, and that they are to be orthogonal (in the
effectively $N=2$ supersymmetric configuration) to the direction $(x^6)$ in 
which the Dirichlet 4-branes are finite and terminate on 5-branes.

The unique low-energy effective action for the (dimensionless) 
$\o$-hypermultiplet in the presence of the FI-term reads \cite{ikz}:
$$ S_{EH}[\o]=-\,\fracmm{1}{2\kappa^2}\int d\zeta^{(-4)} du \left\{
\left(D^{++}\omega\right)^2-\fracmm{(\x^{++})^2}{\o^2}\right\}~,\eqno(13)$$
where $\x^{++}=u^+_iu^+_j\x^{(ij)}$ is the FI-term, and $\kappa$ is the
coupling constant of dimension one (in units of length). After changing the
variables to $q^+_a=u^+_a\omega + u^-_af^{++}$, and eliminating the Lagrange
multiplier $f^{++}$ via its algebraic equation of motion, one can rewrite
eq.~(13) to the equivalent gauge-invariant form
$$ S_{EH}[q,V]=-\,\fracmm{1}{2\kappa^2}
\int d\zeta^{(-4)} du \left\{ q^{a+}_A D^{++}q^+_{aA}+
V^{++}\left(\frac{1}{2}\varepsilon^{AB}q^{a+}_Aq^+_{Ba}+\x^{++}\right)
\right\}~, \eqno(14)$$
in terms of {\it two\/} FS hypermultiplets $q^+_{aA}$ $(A=1,2)$ and the 
auxiliary real analytic $N=2$ vector superfield $V^{++}$~\cite{ikz}, where we
have introduced the pseudo-real notation $q_a=(\sbar{q}{}^+,q^+)$ and
$\ve^{ab}q^+_b=q^{a+},~ a=1,2$. It is
now straightforward to calculate the bosonic terms in the HSS action (14),
in terms of the scalar fields, $\left.q^+_A\right|=f^i_Au^+_i$, and 
$f^i_A\equiv m^i_A\exp(i\vf^i_A)$. One finds the constraint 
$$\x^{(ij)}= \bar{f}_{1}^{(i}f^{j)}_2-f^{(i}_1\bar{f}^{j)}_2~,\eqno(15)$$
leading to the Eguchi-Hanson metric for the kinetic terms, as well as the 
scalar potential \cite{ku}
$$ V=\fracmm{Z\Bar{Z}}{(f_1\bar{f}_1+f_2\bar{f}_2)}
\left[ (f_1^i\bar{f}_{2i}-f_2^i\bar{f}_{1i})^2
+(f_1^i\bar{f}_{1i}+f_2^i\bar{f}_{2i})^2\right]~.\eqno(16)$$
When choosing the direction $\x^{2}= \xi^{3}=0$ and $\xi^{1}=2i$, it is not
difficult to solve the constraint (15) in terms of four independent fields
$ \abs{f_2^1}\equiv  m, \quad \abs{f_2^2}\equiv n, \quad \varphi_1^1 \equiv
 \theta~,\quad \varphi_2^2 \equiv \phi~,$ where the local $U(1)$ invariance has
been fixed by the gauge condition $\varphi_2^1+\varphi_2^2
=\varphi_1^1+\varphi_1^2$. One finds \cite{ku}
$$
V= \fracmm{\abs{Z}^2\sin^2(\theta +\phi)}{m^2+n^2}\left[
\fracmm{4(m^2-n^2)^2}{1+(m^2+n^2)^2\sin^2(\theta +\phi)}
+\fracmm{1+(m^2+n^2)^2\sin^2(\theta +\phi)}{\sin^4(\theta +\phi)}\right]~.
\eqno(17)$$
It is clear that the potential $V$ is positively definite, and it is only 
non-vanishing due to the non-vanishing central charge $\abs{Z}$. It signals 
the spontaneous breaking of $N=2$ supersymmetry in our model.


\begin{thebibliography}{77}

\bibitem{sw} N. Seiberg and E. Witten, {\it Nucl. Phys.} {\bf B426} (1994) 19;
{\bf B431} (1994) 484 
\bibitem{tman} S. Mandelstam, {\it Phys. Rep.} {\bf C23} (1976) 245 \\
G. t'Hooft, {\it Phys. Scr.} {\bf 25} (1982) 133
\bibitem{amz} L. Alvarez-Gaum\'e, M. Marino and F. Zamora, {\it Softly broken
N=2 QCD with massive quark hypermultiplets}, CERN preprints TH/97--37 and 
TH/97--144, hep-th 9703072 and 9707017
\bibitem{wi} A. Hanany and E. Witten, {\it Nucl. Phys.} {\bf B492} (1997) 
152\\
E. Witten, {\it Nucl. Phys.} {\bf B500} (1997) 3
\bibitem{gikos} A. Galperin, E. Ivanov, S. Kalitzin, V. Ogievetsky and
E. Sokatchev, {\it Class. and Quantum Grav.} {\bf 1} (1984) 469
\bibitem{my} S. Ketov, {\it On the exact solutions to quantum N=2 gauge
theories}, DESY and Hannover preprint, DESY 97--199 and ITP-UH-26/97,
hep-th 9710085
\bibitem{next} M. Henningson, {\it Nucl. Phys.} {\bf B458} (1996) 445 \\
M. Matone, {\it Phys. Rev. Lett.} {\bf 78} (1997) 1412 \\
S. Ketov, {\it On the next-to-leading-order correction to the effective 
action in N=2 gauge theories}, DESY and Hannover preprint, DESY 97--103 and
ITP-UH-18/97, hep-th 9706079 \\
J. de Boer, K. Hori, H. Ooguri and Y. Oz, {\it K\"ahler potential and higher
derivative terms from M theory fivebrane}, Berkeley preprint, hep-th 9711143
\bibitem{bko} I. Buchbinder, S. Kuzenko and B. Ovrut, {\it On the D=4, N=2
non-renormalization theorem}, Philadelphia, Princeton and Hannover preprint,
UPR-775T, IASSNS-97/109 and ITP-UH/25/97, hep-th 9710142 
\bibitem{ikz} E. Ivanov, S. Ketov and B. Zupnik, {\it Induced hypermultiplet
self-interactions in N=2 gauge theories}, DESY, Hannover and Dubna preprint,
DESY 97--094, ITP-UH-10/97 and JINR--97--164, hep-th 9706078
\bibitem{ku} S. Ketov and Ch. Unkmeir, {\it Induced scalar potentials for
hypermultiplets}, DESY and Hannover preprint, DESY 97--206 and ITP-UH-28/97,
hep-th 9710185
\end{thebibliography}
\end{document}